\documentclass[submission,copyright,creativecommons]{eptcs}
\providecommand{\event}{FESCA 2016} 

\usepackage{graphicx}
\usepackage{subcaption}
\usepackage{caption}
\usepackage{float}
\usepackage{hyperref}
\hypersetup{
    colorlinks=true,
    citecolor=black,
    linkcolor=black,
    urlcolor=black,
    linktoc=all
}
\usepackage{tabularx}
\usepackage{multirow}
\usepackage{array}
\usepackage{adjustbox}
\usepackage{graphicx}
\usepackage[T1]{fontenc}

\title{Development and Validation of Functional Model of a Cruise Control System}

\author{\quad \quad Avinash Visagan Varadarajan
\institute{\quad \quad Eindhoven University of Technology}
\email{\quad \quad a.v.varadarajan@student.tue.nl}
\and
Marcel Romijn
\institute{{BRACE} Automotive B.V}
\email{\quad marcel.romijn@brace-automotive.com}
\and
Bart Oosthoek
\institute{{BRACE} Automotive B.V}
\email{bart.oosthoek@brace-automotive.com}
\and
Joanna van de Mortel-Fronczak
\institute{Eindhoven University of Technology}
\email{j.m.v.d.mortel@tue.nl}
\and
Jos Beijer
\institute{{HAN} University of Applied Sciences}
\email{jos2076@hotmail.com}
}

\def\titlerunning{Functional Model of Cruise Control}
\def\authorrunning{A. V. Varadarajan, M. Romijn, B. Oosthoek, \\ \ J. v. d. Mortel-Fronczak \& J. Beijer}

\begin{document}

\maketitle

\begin{abstract}
Modern automobiles can be considered as a collection of many subsystems working with each other to realize safe transportation of the occupants. Innovative technologies that make transportation easier are increasingly incorporated into the automobile in the form of functionalities. These new functionalities in turn increase the complexity of the system framework present and traceability is lost or becomes very tricky in the process. This hugely impacts the development phase of an automobile, in which, the safety and reliability of the automobile design should be ensured. Hence, there is a need to ensure operational safety of the vehicles while adding new functionalities to the vehicle. To address this issue, functional models of such systems are created and analysed. The main purpose of developing a functional model is to improve the traceability and reusability of a system which reduces development time and cost. Operational safety of the system is ensured by analysing the system with respect to random and systematic failures and including safety mechanism to prevent such failures. This paper discusses the development and validation of a functional model of a conventional cruise control system in a passenger vehicle based on the ISO 26262 Road Vehicles - Functional Safety standard. A methodology for creating functional architectures and an architecture of a cruise control system developed using the methodology are presented. 
\end{abstract}

\section{Introduction}

The complexity of a modern automobile is increasing by the day and the development of such a vehicle is a challenge by itself. Furthermore, the demands of the customers with respect to additional functionalities indirectly increase the complexity of the design. This is caused by the fact that the functionalities are incorporated in a highly interactive environment of Electrical and Electronic (E/E) systems which adds further to the interdependencies and interactions between them. In such modern cars, the failure of an E/E system can cause the loss of control or misbehaviour of the vehicle that can result in fatal crashes. This also results in life threatening injuries to the driver and other motorists. Hence, ensuring operational safety is of paramount importance in these vehicles. However, adding new functionalities and validating the operational safety of a new subsystem in the existing system framework increases development time.  Thus the reconsideration of fundamental electronics in terms of functional behaviour will be highly beneficial in terms of reducing complexity and improving operational safety as proposed in \cite{bergmiller2013design}. This will not only support the development of innovative functionalities, but also make them safer without increasing the development time. According to International Electrotechnical Commission \cite{ieccite}, functional safety can be defined as the detection of a potentially dangerous event and activation of a corrective measure to prevent or mitigate the consequence of the event. According to ISO 26262 Road Vehicle - Functional Safety standard \cite{iso201126262}, functional safety can be defined as the avoidance of risks caused by systematic failures and random hardware failures. These failures are a result of increasing technological complexity, software content and mechatronic implementation.

\paragraph{}

The {ISO} 26262 standard provides a number of methods and processes to ensure functional safety. Problems related to speed of development and compatibility arise when different Original Equipment Manufacturers ({OEM}) or even different departments within the {OEM} use different methods to ensure functional safety. For example, if some results of the Hazard Analysis and Risk Assessment ({HARA}) process are validated using a comprehensive Simulink model, the actual vehicle test becomes less expensive. On the other hand, the results from exhaustive vehicle testing can be implemented in a relatively simple Simulink model. Both methods produce {ISO} 26262 standard compliant systems. Such inconsistencies and differences will create problems in subsequent development stages. To tackle this problem, this paper proposes the following mentioned methodology to facilitate efficient functional model development while being fully compliant with the {ISO} 26262 standard. This breakdown methodology based on the {ISO} 26262 standard also ensures hierarchy and traceability.  The approach followed in developing a functional model of the cruise control system is as follows:
\newline
\begin{enumerate}
\item Identify the components and interfaces with other subsystems required for the proper functioning of the cruise control and express them as Systems Modelling Language {(SysML)} diagrams.
\item Formulate malfunctions in the system that might cause hazardous behaviour.
\item Perform the hazard analysis and risk assessment ({HARA}) process and assign an Automotive Safety Integrity Level ({ASIL}) rating for each hazard. 
\item Validate the result of the {HARA} process with a Simulink model of the crusie control system.
\item Define a safety goal to prevent each hazard or mitigate the effect of it in case the hazardous event is not preventable. 
\item Consequently, derive functional safety requirements for each defined safety goal.
\end{enumerate}

\paragraph{}
In this paper, we show how a functional model of an E/E system can be developed and validated. A functional model is a representation of the system that concerns its input and output behaviour. Building and analyzing a functional model of an E/E system will improve the perception of the vehicle behaviour analogous to proper functioning or malfunctioning of the system in consideration. This is possible as the system behaviour is studied in response to various erroneous inputs.  The output of the system is studied by the process of `Fault Injection' to various `Fault Entry' points in the functional model. These fault injection points are parts of the cruise control model representing malfunctions which cause risky behaviour of the vehicle. The effects of fault injection can be analysed with the help of the Simulink model of the cruise control system. Some examples of fault injection points are: erroneous speed sensor data, faulty inputs from driver interface or improper actuation of the throttle valve. If unexpected system behaviour is encountered, functional safety requirements that prevent the related hazardous events or mitigate their effects are formulated. These requirements are then included in the actual system design. The definitiveness of this functional model will be validated with the help of the Simulink model that closely resembles the behaviour of the corresponding real-life system. To support the description of this process, a case study of an E/E system is presented. The E/E system taken into consideration is the conventional cruise control system. The functional model of this system will be built hierarchically based on the ISO 26262 standard safety lifecycle \cite{iso201126262} and the process will offer seamless traceability to the components affected due to fault injection. Although the standard provides a fixed set of guidelines for the development of a model of E/E system in the concept phase, it has a wide scope for development as it does not prescribe a fixed methodology to achieve a risk-free/risk-reduced system that satisfies the guidelines. For example, the objective of item definition according to the {ISO} 26262 standard is that the item under consideration has to be defined and described with respect to its functionality, interactions with other systems and the environment. But it does not specify clearly the methods, techniques or tools that can be used for this purpose. 

\paragraph{}

The paper is organised as follows. Section \ref{relw} presents results from related work. In Section \ref{fmccs}, the step-by-step development of the functional model of a cruise control system is explained and the important functions for the proper functioning of the cruise control system are identified. Deviations from these important functionalities are used to identify potential hazardous situations. The hazard analysis and risk assessment procedure is explained in Section \ref{harap}. Section \ref{valid} explains the validation of the functional model. In Section \ref{fsc}, the functional safety requirements which ensure functional safety of the system are derived from the safety goals. Finally, the conclusions are presented in Section \ref{conc}.


\section{Related Work} \label{relw}
Since the majority of concepts is based on the ISO 26262 standard, this overview is mainly focused on parts of the standard such as item definition, HARA and functional safety concept. To start with, \cite{kannananalysis} pairs the different safety techniques used in the automotive domain to the sections of the ISO 26262 standard. It provides a distribution of safety techniques, where they are divided based on the sections of ISO standards which they addressed.
Functional testing of an E/E system with the help of a Simulink model is vital in identifying the correct functioning of the system. Such a testing procedure should also cover the malfunctions proposed by the HARA process. \cite{hara1} explains the hazard identification using various techniques like Fault Tree Analysis (FTA), Failure Modes Effects Analysis (FMEA) and Common Cause Analysis (CCA). The process of identifying a hazard using FMEA technique is of interest to this paper. To achieve this, the concept of fault injection has been introduced. Formulation of an effective safety concept for a hazard in a hybrid vehicle is explained in \cite{harafsc}. Various other papers such as \cite{sahara}, \cite{wardgenericapproach} and \cite{mbha} provide a detailed methodology for the identification and classification of hazards. However, as pointed out by \cite{rana2013increasing}, much of the testing is done in the verification and validation phase and the usage of fault injection on a model level will lead to a development  of a correct system right from the initial phases. Functional architectures are provided in  \cite{lotz2013system}, \cite{geyer2013maneuver}, \cite{bergmiller2013design} for systems such as steer-by-wire, automated driving and various drive-by-wire systems in a vehicle. Moreover,  \cite{bergmiller2013design} also explains in detail the development of a steer-by-wire system including functional requirements and functional architecture on system and sub-system levels. \cite{fscg} and \cite{fsr1} deal with formulating functional safety requirements from which functional safety architectures are developed. Another paper worth mentioning is \cite{mbsefs} which describes the development of an automotive functional safety case in SysML and Goal Structured Notation environments. However, there seems to be a lack of a common framework for designing the architecture of any E/E system and nearly all the architectures found in literature correspond to one particular system. The functional architecture proposed in this paper can be used for functional validation and verification of almost all E/E systems. It can be seen from literature review that many works have concentrated on improving specific parts of the ISO standard like requirement modelling, hazard analysis, risk assessment and formulation of safety concepts for identified hazards. 

\paragraph{}


\section{Functional model of cruise control system} \label{fmccs}
In order to construct a functionally safe system, it is important to design a well-decomposed functional architecture of the system. This decomposed architecture helps in analysing the localised effects of malfunctions in different components. It also provides flexibility in implementing the safety mechanism that reduces the impact of the malfunctions.  With respect to the cruise control model, it will be decomposed into vehicle level, system level and subsystem level abstractions. A decomposed functional model enables fault injection and analysis of its effects at the above mentioned abstraction levels. In addition to defining the structure, the behaviour of the system should also be defined.

\begin{figure}[h]
\centering
\includegraphics[width=13cm,height=13cm,keepaspectratio]{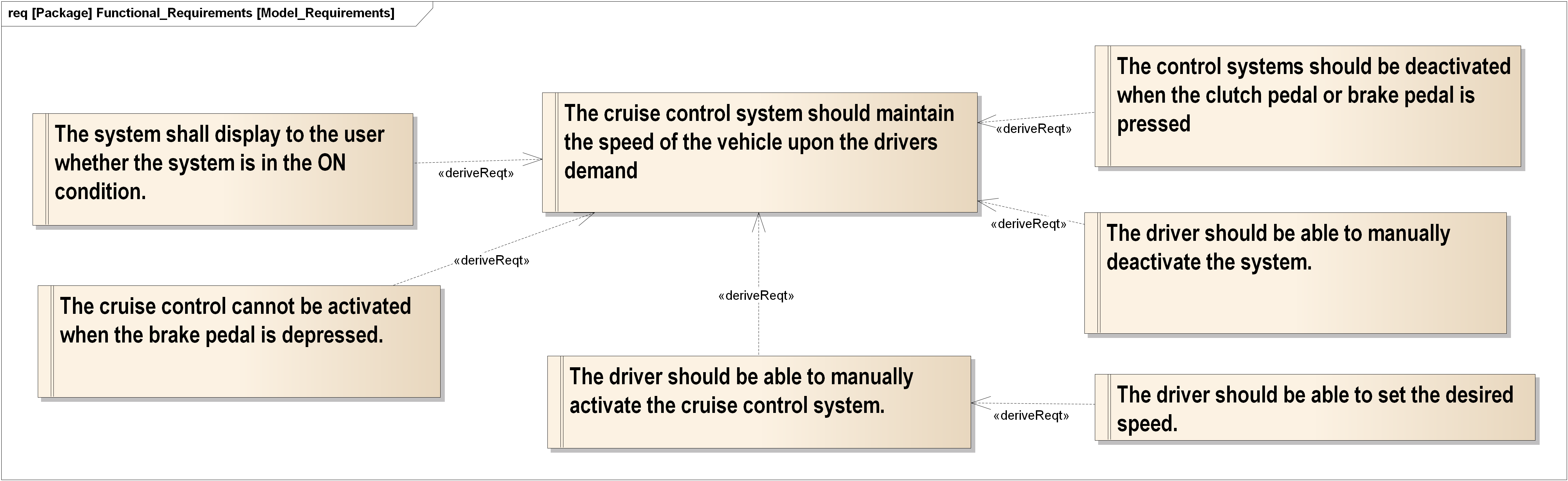}
\caption{Functional requirements of the cruise control}
\label{fig:mr}
\end{figure}

\paragraph{}
The cruise control system maintains the speed of the vehicle according to the input given by the driver. For simplifying the analysis, only the longitudinal motion of the vehicle is considered. During the operation of the cruise control, pressing the clutch or brake pedal will automatically deactivate the cruise control. However, the driver is allowed to accelerate using the accelerator pedal when the cruise control is in operation. It is important to note that the cruise control will not manipulate the brake system of the vehicle. This means that when the vehicle with the cruise control system approaches another vehicle, it will not automatically slow down. At the start of the development process, the above mentioned functionalities of the cruise control system are represented in a {SysML} requirement diagram as given in Figure \ref{fig:mr}. The figure shows the derivation of various sub-functionalities of the cruise control system from its main functionality in the form of requirements.

\begin{figure}[h]
\centering
\includegraphics[width=15cm,height=8cm,keepaspectratio]{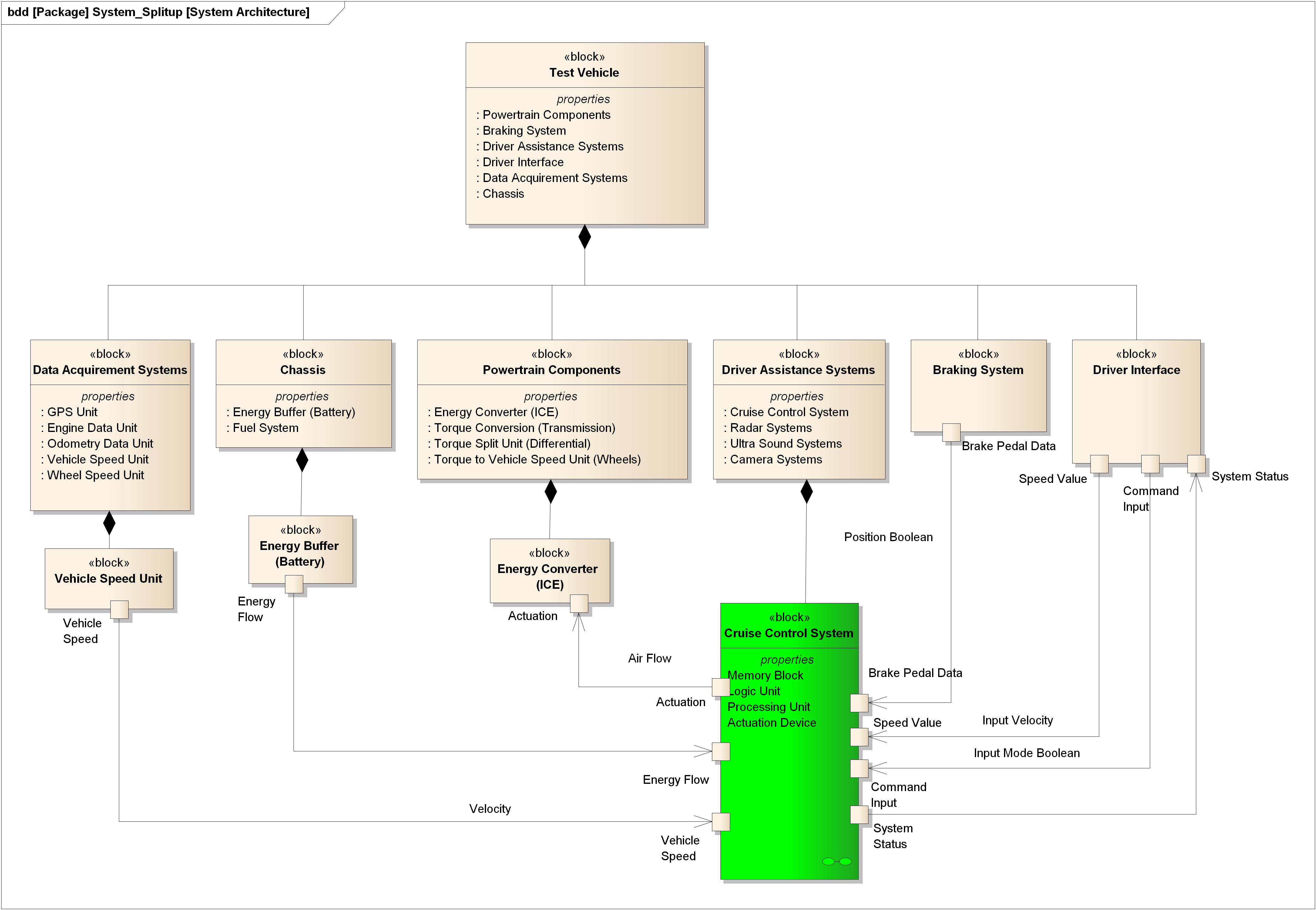}
\caption{System level architecture of the cruise control system}
\label{fig:sl}
\end{figure}

\begin{figure}[]
\centering
\includegraphics[width=13cm,height=13cm,keepaspectratio]{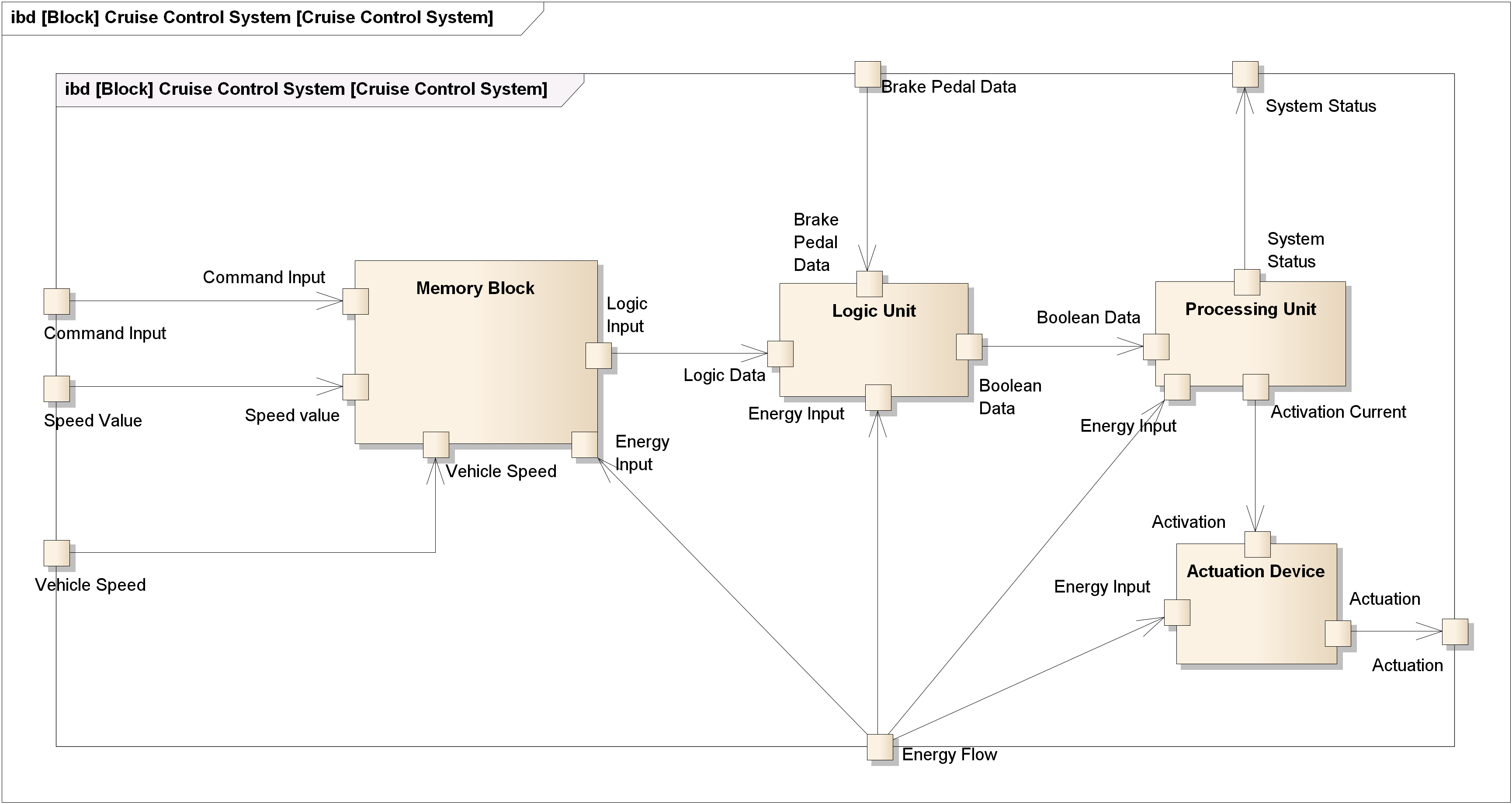}
\caption{Subsystem level architecture of the cruise control}
\label{fig:sslarc}
\end{figure}

The formulated cruise control requirements are translated into functional components in Figures \ref{fig:sl} and \ref{fig:sslarc}. Each translated functional component should at least be coupled to one functional requirement to ensure traceability. The coupling is important, as any malfunction can be traced back to the respective requirement and ultimately, it can be found which functionality is breached. A vehicle-level architecture represents the important components of the vehicle to successfully achieve the intended functions of the system. The inputs from the user are translated to vehicle functions through the actuators. The sensors provide feedback of vehicle velocity to the cruise control system to ensure proper working. In essence, the cruise control system is considered as a blackbox that receives input from the user and maintains the speed of the vehicle.
\paragraph{}

The system-level architecture is the representation of the cruise control system in a system-level environment within the vehicle. It includes the boundaries of the system, including the important interaction with other components. Figure \ref{fig:sl} represents the system-level architecture. The figure represents the cruise control system as a black-box and provides the interfaces to the components of the vehicle required for its proper functioning. It clearly shows the inputs obtained from the driver interface, vehicle speed sensor and braking system and the output provided by it to control the throttle position. The vehicle data acquisition system collects all the required sensor data such as vehicle speed, wheel speed and engine output torque. These are the major driver independent inputs required for the cruise control system. One of the major vehicle components is the powertrain system, which helps in the transmission of power to the wheels. In case of an unavoidable emergency situation during the operation of the vehicle caused by malfunction of the cruise control system, the control of powertrain system will be deactivated to prevent harm to the driver. The system receives input from the driver interface and provides the control output in the form of actuation signal to the internal combustion engine. The signals from the driver interface, when the driver activates/de-activates, sets the speed to be maintained, puts the system on standby or resumes the cruise control, are transmitted as input signals to the cruise control unit. The cruise control system also obtains data from the transmission system regarding the clutch position and from the braking system it obtains the brake pedal position. Some physical flows like the flow of energy from the energy buffer (battery), have been modelled as well. The status of the cruise control is displayed to the driver as feedback through the driver interface. In modelling the cruise control system environment, it is assumed that all other related systems such as steering function, braking function, transmission, power supply system, etc. are working perfectly. 

In the subsystem-level representation of the cruise control, the blocks have been organized as Input, Output and sections. The subsystem-level architecture is presented in Figure \ref{fig:sslarc}. The figure provides a pseudo-architecture of the internal organisation of the cruise control block. This pseudo-architecture generalises the cruise control components as the input, logic and the output blocks. The memory block represents the input block, where the input from the driver is stored and given for processing. The input can vary from on/off commands of the cruise control to the analogue speed values set by the driver. The logic unit represents the logic block. The inputs are processed for their logical correctness in the logic unit. The main function of the logic unit is to check for input redundancy and prevent misuse of the cruise control. The processing unit represents the output block. The processing unit receives input from the logic unit and gives control input in the form of activation current to the actuation device. After the definition of functional safety concept in Section \ref{fsc}, the processing unit will be responsible for choosing between the signal generated based on driver input or the safe state signal generated in case of unexpected behaviour caused by cruise control system. The actuation device is used to control the output power from the energy converter (IC Engine) to the wheels. The block definition diagram (bdd) and the internal block diagram (ibd) represent the technical components of the vehicle as pure functional blocks including their input/output behaviour on a functional level. Creating such a functional model is advantageous from an engineering perspective, as it gives freedom while selecting the technical components to satisfy the functions and also allows reusability of the models while developing similar systems. The functional models are also hugely advantageous to study interactions with other related and resource-sharing systems. Thus the functional models can be reused multiple times in model creation by making very small changes according to the system in consideration respecting the change management requirements of the {ISO} 26262 standard. This saves substantial time during product development and testing and thus reduces the development cost, all while strictly following the guidelines proposed by the ISO 26262 standard. 

\begin{figure}[h]
	\centering
    \includegraphics[width=13cm,keepaspectratio]{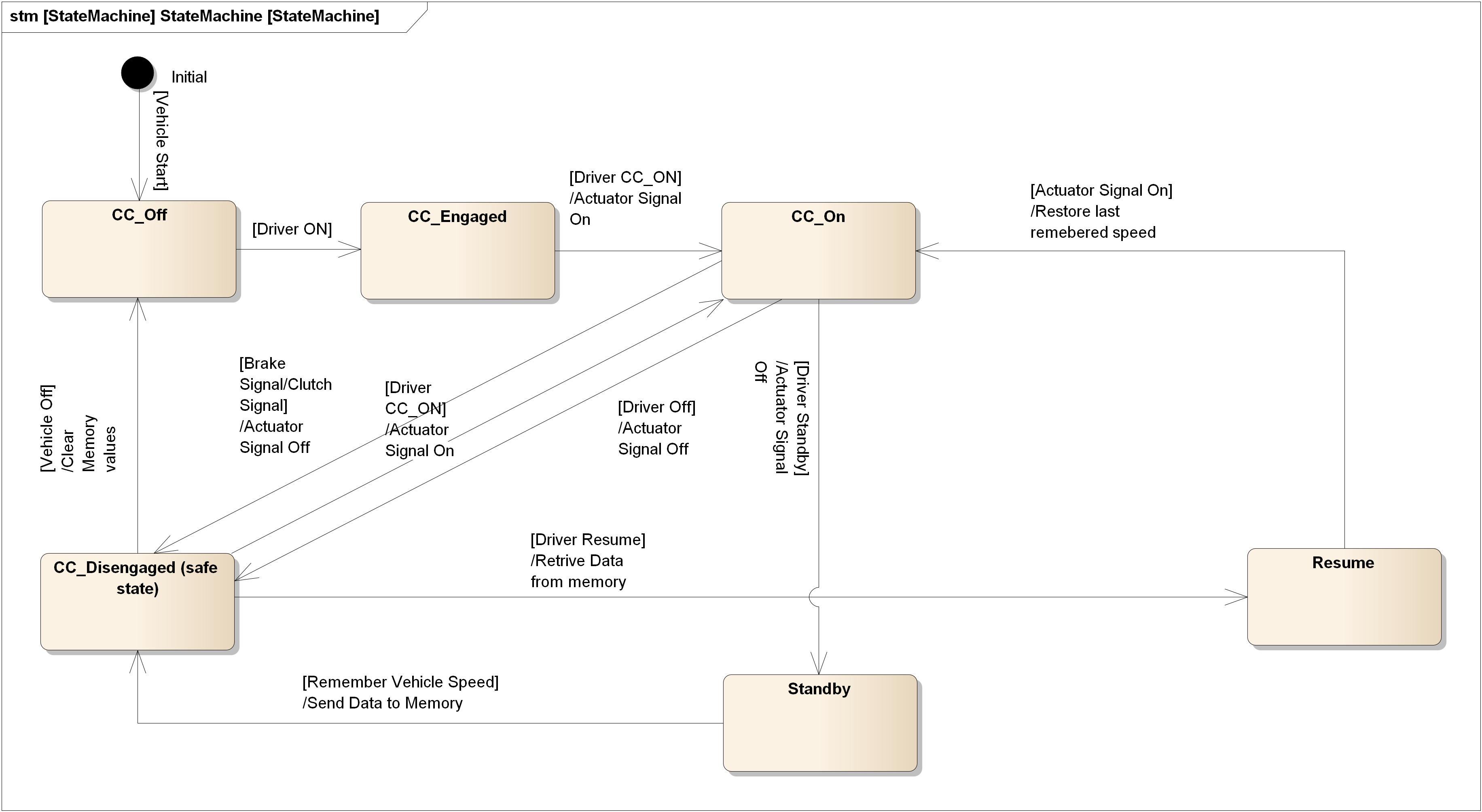}
	\caption{Functional state transitions of the Cruise Control}
	\label {fig:stmfreq}
\end{figure}

It is important to represent the transitions among the states of the cruise control in addtion to the architecture. These transitions can also be interpreted as the behaviour of the system in response to inputs from the user or the sensor. To represent the behaviour, the transition between the states of the cruise control can be modelled with the help of {SysML} state machine diagrams (stm). As explained in \cite{delligatti2013sysml}, the stm acts as an input to development  and it is suitable for the behaviour representation of cruise control as the system has defined states. Figure \ref{fig:stmfreq} graphically describes the transitions between the states corresponding to certain triggers or inputs from the driver or adjacent systems. The blocks in the figure represent the possible states of the cruise control system and the arrows represent the triggers or inputs which lead to the transition from one state to another. Some of the states may be automatically attained in case of inadvertent vehicle behaviour triggers while other are attained through systematic inputs. Initially, the cruise control is in the Off state when the vehicle is started. Only from an input through the driver interface (Driver On), the cruise control system can transition to the Engaged state. At this point, the driver needs to give the set speed value to the system and only when the cruise control system gives out signal to the actuator (Driver CC\_On / Actuator Signal On), the cruise control is in the On state and the driver is notified through the driver interface. From the on state, it is possible to go to the Disengaged state by either pressing the clutch, pressing the brake pedal or from the driver interface (Brake Signal/ Clutch Signal / Actuator Signal Off) (Driver Off / Actuator Signal Off). It is also possible to put the cruise control in the Standby mode, in which case it transitions to the Disengaged state (Driver Standby / Actuator Signal Off). In any case, the cruise control should remember the last value of speed maintained by the system. From the Disengaged state, the driver can either resume the operation of the cruise control or start the operation all over again by pressing the cruise control button (Driver CC\_On / Actuator Signal On). If no action is taken, the cruise control remains idle and when the vehicle is switched off, the cruise control system is also cutoff from the power supply and the existing values in the cruise control memory are erased (Vehicle Off / Clear Memory values). The Disengaged state is considered as a safe state because the cruise control system stops controlling the actuator and returns the throttle control of the vehicle back to the driver. From this breakdown approach, the important vehicle-level functionalities related to the cruise control are identified. When the system deviates from these functions, the system is said to be malfunctioning. The malfunctions are the cause of hazardous situations and act as input for the subsequent hazard analysis and risk assessment process. The following list provides those functionalities.

\begin{enumerate}
\item Deliver requested power to the wheels.
\item Acquire the current speed of the vehicle and provides the speed value to the processing unit.
\item Provide signal to the processing unit when the brake pedal/clutch pedal is pressed.
\item Provide input from the driver to the processing system.
\item Provide electrical energy to connected components.
\end{enumerate}


\section{Hazard analysis and risk assessment} \label{harap}
\begin{figure}[h]
	\centering
    \includegraphics[width=10cm,keepaspectratio]{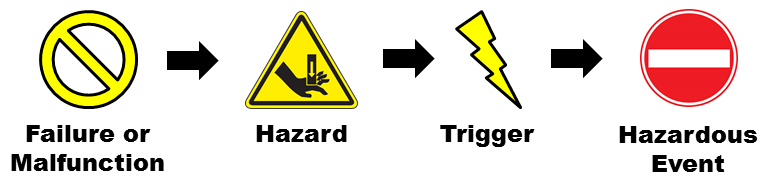}
	\caption{Hazardous event formation}
	\label {fig:hazf}
\end{figure}

In order to identify hazards, a list of guidewords as defined in \cite{surfacevehicleprac} is used to derive the failures caused by deviation from normal behaviour. The guidewords are loss of function, more function than intended, less function than intended, wrong direction of function, unintended activation of function and failure of function to update as intended. The relevant guidewords based on the context to the functionalities listed above are selected and applied, giving rise to the failure description. Based on the failure description, the hazard is identified. After identification of the hazards, hazardous events are defined for relevant combinations of operational situations and the identified hazards. The operating situations are defined in terms of use cases. The use cases are nothing but a formal representation of interaction between the system and actions or triggers that lead to hazardous events. The use cases also specify the operating environment of the vehicle in which the system is used (for example, highway, off-road conditions, snow and ice).  Based on hazard and the situation, the consequences to the occupants and the other concerned road users are determined. Figure \ref{fig:hazf} graphically shows how a hazardous event arises from a failure or malfunction. Such hazardous events cause risks to passengers or other road users. The subsequent list identifies the risks to the passengers or other road users caused by the hazards.

\begin{enumerate}
\item Accident due to unintended acceleration.
\item Accident due to unintended deceleration.
\item Accident due to unintended movement (acceleration or deceleration due to late/sudden activation of the cruise control).
\item Accident due to reduced movement (engine braking).
\item Accident due to dangerous movement (jerky movement of the vehicle due to fluctuating actuation).
\item Accident due to unintended reaction (unexpected reaction of the CC system due to CAN message errors).
\end{enumerate} 

\paragraph{}

Hazardous events are created by the combination of use cases and the potential failures. This results from a matrix of hazards occurring at a particular situation. The situation analysis has to be performed first. It gives an overview of the hazard causing situations in terms of three main parameters namely \emph{Operating situation, Environment situation, Traffic Situation}. To make a more in-depth situation analysis, the important operating modes of the vehicle [\emph{Medium Speed, High Speed, Cornering}] are also included.  The hazardous events are classified based on their severity, exposure and the controllability. The use cases defined earlier also play a significant role in determination of the exposure. For example, a vehicle with a cruise control system is driven substantially more on a highway when compared to off-road conditions. Hence the exposure to the first use case is higher compared to the second). From the individual severity, exposure and controllability ratings, the Automotive Safety Integrity Level ({ASIL}) rating is assigned to the hazardous situation from the matrix available in Section 7.4.4 of ISO 26262-3 as defined in \cite{iso201126262}. The hazardous events can have any of the {ASIL} ratings of {QM}, A, B, C, D where {QM} is the lowest integrity level and D is the highest integrity level. As the integrity level increases, the danger of the hazard also increases. For every hazardous event defined, a safety goal should be created. The safety goal should be integrated in the functional architecture and this process results in making the functional safety architecture ({FSA}).

\paragraph{}
The next step is the formulation of safety goals. They are the top-level safety requirements of an automobile. They should be relatable to the vehicle-level architecture. They should be formulated on a functional level without much technical representations. A safety goal should be formulated for every hazardous event identified. The safety goal shall be assigned an {ASIL} rating corresponding to the highest {ASIL} rating determined from operational situations. The safety goal also recommends transition to a safe state according to the state machine diagram of the {CC} system as shown in Figure \ref{fig:stmfreq}. The Hazard Analysis and Risk Assessment ({HARA}) procedure is pivotal in shaping the components required for satisfying the operational safety of the system. The safety goals are the outcome of the {HARA} procedure from which the functional safety requirements are derived. The {HARA} procedure does not include granulated analysis of operational situations as it might potentially lead to a reduction of {ASIL} ratings. From the safety goal determination, it is also important to assign safe states to the safety goals, if a safe state is available in the states mentioned in item definition. In the case of the cruise control system, a `CC\_ Disengaged' state is available. This particular state is considered as the cruise control system will not be able to control the actuator and so any malfunctions in the cruise control system will not affect the vehicle behaviour. The main aim of introducing a safe state is that, when a hazard is identified by the system, the system transitions to the safe state. In this way, the risk caused by the hazards will be reduced. Table \ref{tab:sg} provides the safety goals of the cruise control system.  

\begin{table}[]
\resizebox{\textwidth}{!}{
\tiny
\centering
\begin{tabular}{| >{\centering\arraybackslash}m{1cm}| >{\centering\arraybackslash}m{5cm}| >{\centering\arraybackslash}m{1cm}| >{\centering\arraybackslash}m{2cm}|}
\hline
\textbf{Safety Goal ID} & \textbf{Safety Goal Description} & \textbf{ASIL Rating} & \textbf{Safe State} \\ \hline
SG01 & The system should not transfer excess power to the wheels where it causes unintended acceleration & C & CC\_ Disengaged \\ \hline
SG02 & The system should not reduce the supplied power to the wheels where it causes unintended deceleration & C & CC\_ Disengaged \\ \hline
SG03 & The system should not allow a sudden loss in power to the wheels where it causes unintended reduction in movement & QM & CC\_ Disengaged \\ \hline
\end{tabular}}
\caption{Safety goals}
\label{tab:sg}
\end{table}

In the {HARA} process and ASIL determination given above, it is found that in some cases the {ASIL} rating is a C. This is mostly due to high exposure rating. The high exposure rating was caused by the high `probability' of the system being exposed to a highway or a city road. Besides considering the probability of exposure, it is also important to consider the occurrence frequency of the hazard. The higher {ASIL} rating is not entirely due to the malfunction of the system. The hazardous events are caused by certain triggers for which the frequency of occurrence is very low while the probability of the exposure to the operating situation might be very high. For example, a vehicle with a cruise control system might be driven on a city road frequently. But the frequency of exposure of a malfunctioning system at high speeds on a city road is quite low, as there are constraints on the speed limits and the presence of corners. The high {ASIL} rating of one such operating situation will contribute to a very high {ASIL} rating of the hazardous event. This results in increased costs for safety tests and in developing an added layer of security that will increase the complexity of the system and the development time. Therefore, care has to be taken to ensure that the ratings are correct and unnecessarily high ratings should be avoided.   


\section{Validation of the functional model} \label{valid}
This section is concerned with validation of the effects of the hazards which result from the hazard identification of the {HARA} process and the {ASIL} ratings. For the validation, we consider fault injection into the speed feedback.  Two cases are studied where the Simulink model of the cruise control system is fed with i) a lower vehicle speed, ii) a higher vehicle speed feedback. A higher {ASIL} rating during the development phase demands strict and thorough testing once the system is designed. The validation of the functional model also indicates if the {ASIL} rating assigned  to a hazardous event is appropriate. In case the simulations prove that the rating assigned from the {HARA} is indeed higher than the actual risk rating, there might be a recommendation to lower the {ASIL} rating. This will reduce the time devoted to testing the effects due to the particular hazard. The validation is done by the introduction of `Fault Injection' points at various points in the model. These fault injection points model the malfunctions introduced in the previous section. It is important to note that cornering behaviour and certain environmental conditions are difficult to model accurately in the Simulink model and the simulation is restricted to motion in the longitudinal axis, dry road conditions and medium speed conditions.

\begin{figure}[htp]
\centering
\includegraphics[width=.3\textwidth]{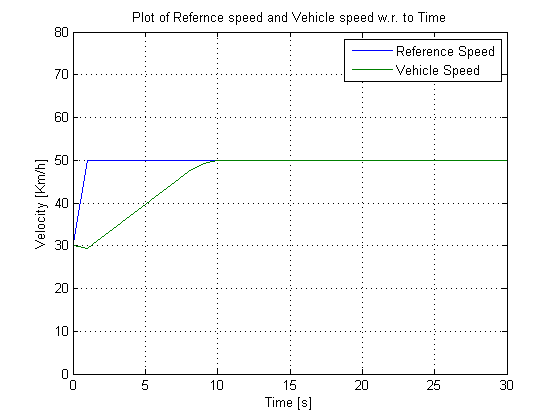}\hfill
\includegraphics[width=.3\textwidth]{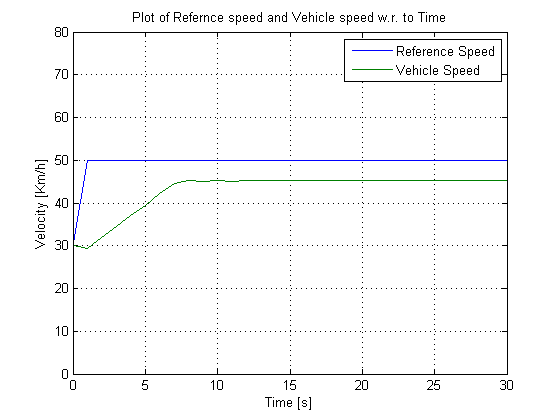}\hfill
\includegraphics[width=.3\textwidth]{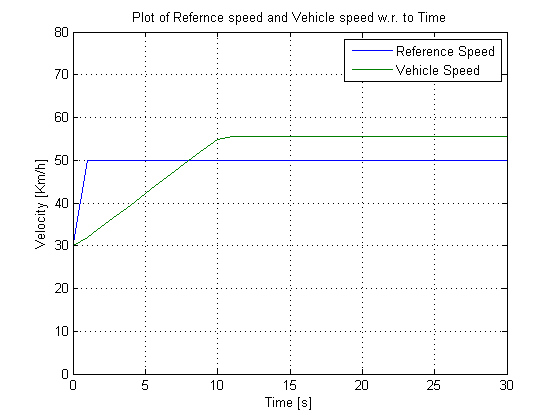}  
\caption{Effects of fault injection - Optimal behaviour (left), Behaviour due to high vehicle speed feedback error (middle), Behaviour due to low vehicle speed feedback error (right)}
\label{fig:fi}
\end{figure}

Figure \ref{fig:fi} represents the results from the Simulink model of the cruise control system. The results are provided in the form of comparison between the optimal behaviour and the actual behaviour of the cruise control due to fault injection. From Figure \ref{fig:fi}, it can be seen that, the cruise control system produces unintended acceleration in case of a lower speed feedback value. Similarly, it produces reduced movement in case of a higher speed feedback. Using this fault injection method, it can be studied whether the designed architecture is functionally safe. This is just one example of the fault injection method. The same method can also be used to study the effects of interference between two such E/E systems. This proves that even though the cruise control system does not malfunction by itself, it performs in an undesirable way due to inappropriate input. One major advantage of validating a functional model is that it provides complete understanding of the effects of the failure at different points of the vehicle.

\section{Derivation of functionally safe architecture} \label{fsc}

\begin{figure}[h]
	\centering
    \includegraphics[width=13cm,keepaspectratio]{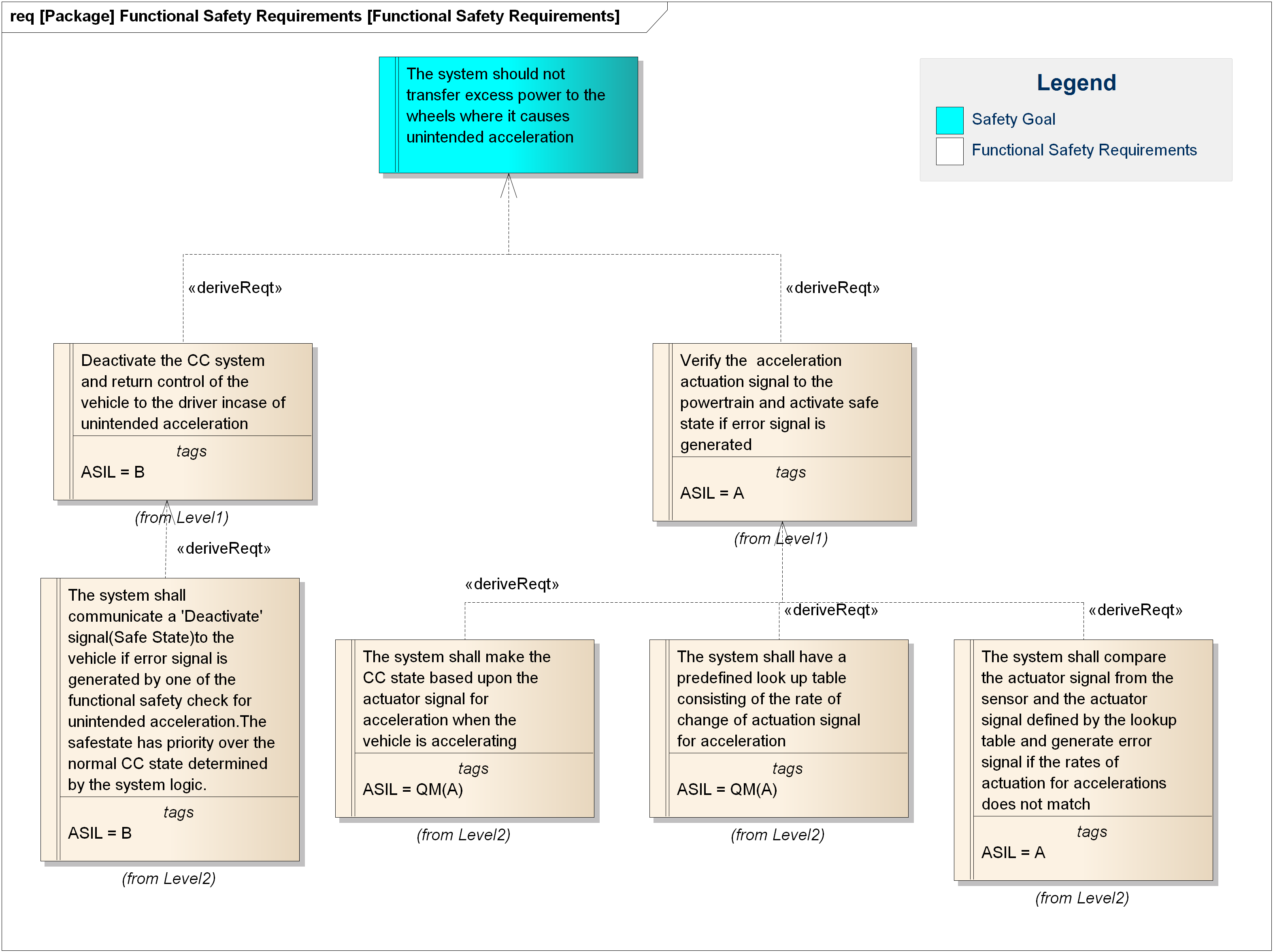}
	\caption{Functional safety requirements to prevent unintended acceleration}
	\label {fig:fsr}
\end{figure}

This section deals with functional safety concept. The main aim of the functional safety concept is the generation of the functional safety requirements and development of the functional safety architecture from the requirements. The safety functions which are represented through the functional safety requirements are implemented in the sub-system layer of the cruise control. At the end of the functional safety concept, a safety functional architecture of the cruise control system with fault detection mechanism are presented. The safety goals from the {HARA} process are fulfilled using the functional safety requirements. The functional safety requirements regarding the cruise control are split into two levels. Level 1 of the {FSR} provides an abstract safety function that addresses the failing function and the reaction from the cruise control system. Level 2 {FSR} is defined in terms of functional \emph{Input, Logic and Output} blocks inside the sub-system architecture. The functional safety requirements which address the hazard `Unintended Acceleration' are given in Figure \ref{fig:fsr}. 

\paragraph{}
Level 1 requirements specify the detection of an actual failure happening and returning the control of the vehicle to the driver in case a failure is detected. From those high-level safety requirements, the level 2 requirements are derived. Going by the pseudo architecture method (generalisation of functional architecture in terms of input, logic and output blocks). The input for failure detection will be the actuator signal. A predefined logic in the form of lookup tables will be made available for comparing the actual signal and the predefined signal. Both of the signals will be compared and if they do not match an error signal will be generated. In case an error signal is generated, preference will be given to the error signal above the normal signal of the cruise control function. When an error signal is generated, the cruise control will be deactivated and the throttle control will be cut off from it. The control of the vehicle will be returned to the driver. Figure \ref{fig:fsr} represents the safety mechanism implemented in the subsystem-level architecture of the cruise control system. The architecture including the safety mechanism can be considered as a functionally safe architecture.

\begin{figure}[h]
	\centering
    \includegraphics[width=13cm,keepaspectratio]{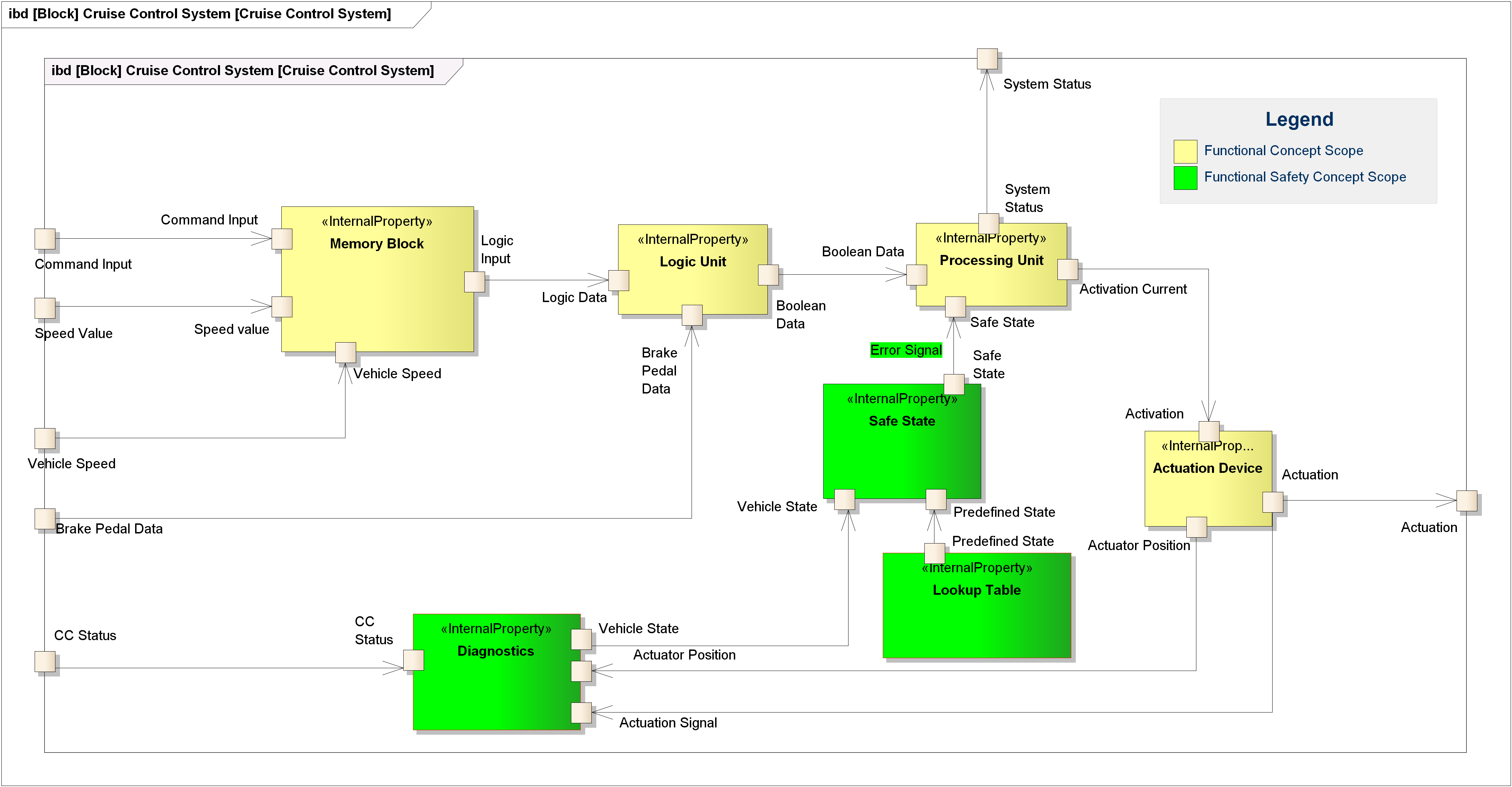}
	\caption{Functional architecture including safety requirements}
	\label {fig:fsca}
\end{figure}

\paragraph{}
The functional safety mechanism of the cruise control system works conceptually as follows. Whenever the cruise control is in the active state, the rate of actuator signal should be monitored and compared with a predefined lookup table which contains the actuation rate for an array of starting speeds and target set speeds. Whenever there is a considerable deviation of the actual rate from the pre-defined rate of actuation, an error signal should be generated. Whenever an error signal is generated, the cruise control should be deactivated and the control of the vehicle should be given back to the driver. Care should be taken to maintain the requirements on a functional level without specifying technical details. Creation of a lookup table and actual comparison is however out of scope of the functional concept. During the actual development of the cruise control system, this safety mechanism must be implemented in the electronic control unit.

\section{Concluding remarks} \label{conc}
The approach followed in this paper is targeted towards functional models which significantly influence organised development of the architecture of cruise control system, as an illustrative case study. The model-based approach followed according to the guidelines of {ISO} 26262 standard provides us with a consistent design of the system with a high level of traceability. Since the faults at any level of the system hierarchy can be traced back to their components and functionality, the process of diagnostics is simplified. In addition, a formal functional architecture has been defined. Upon this architecture, any technical system can be developed and verified with suitable changes. Whenever a functionality is added or changed, only the technical components built upon the functional level have to be changed and the functional concept does not need to be designed again as opposed to a component specific design. This methodology eliminates the need for developing a separate functional architecture, whenever a new system is developed and tested. This results in the reduction of development time and costs. Furthermore, the validation of results of {HARA} process with the help of a Simulink model shows localised effects on the related safety components. This can help the engineers to detect higher {ASIL} ratings at the early stages of development, which will save a lot of time when designing software and hardware components. These results can also be used to make the design robust and capable of withstanding unexpected hardware or software failures. Due to their highly decomposed nature, the functional models allow engineers to analyse and rectify failure of functionality rather than failure of a particular technical component that offers the same functionality. The following points can be considered for improving the capability of functional models for future research.

\begin{enumerate}
\item Separate tools are required for architecture modelling {(SysML)} and validation of the models {(Simulink)}. There does not exist user friendly ways to link both tools at the moment. This can be further researched.
\item In this paper only one system, namely the cruise control system has been considered. Functional architectures for multiple systems can be built and analysed to study the effects of interference of safety functions of one system on another system and the risks caused by them.   
\end{enumerate}


\nocite{*}
\bibliographystyle{eptcs}
\bibliography{bib}

\end{document}